\documentclass[11pt, a4paper]{article}
\usepackage{amsmath, graphicx}

\begin{document}

\author{V. V. Prosentsov\thanks{%
e-mail: prosentsov@yahoo.com} \\
Stationsstraat 86, 5751 HH, Deurne, The Netherlands}

\title{Light amplification and scattering by clusters made of small active
particles: the local perturbation approach}

\maketitle

\begin{abstract}
The light amplification by finite active media is used extensively in modern
optics applications. In this paper, the light amplification and scattering
by the cluster of small active particles is studied analytically and
numerically with the help of the local perturbation method and
phenomenological laser theory. It is shown that light amplification is
possible even for one small particle, and that the amplification is more
profound when the light frequency nears the frequency of the cluster's
morphological resonance. Theoretical discussions are supplemented by
numerical results for scattering by clusters which particles positioned at
ordered and at slightly disordered positions.
\end{abstract}


\section{Introduction}

Light amplification by stimulated emission of radiation (laser) is well
known phenomenon routinely employed in solids-state, gas, and dye lasers
\cite{Shimoda}-\cite{Haken}. With the development of micro and nano
technology the new kind of lasers, so-called photonic crystal lasers,
emerged \cite{Limpert}-\cite{Noda}. The common peculiarity of the photonic
crystal lasers is that they are made of finite number of particles or cells
ordered at least in one dimension. These lasers are extremely versatile:
they can be made from active host medium filled by passive scattering
particles, or from active scatterers immersed into passive host medium, or
both. Moreover, the particles in such lasers can move or be static. From
practical point of view, the photonic crystal lasers and amplifiers can be
used as light sources to compensate the optical losses in metamaterials \cite
{Xiao}-\cite{Wuestner}.

To the author's knowledge, the light generation by a scattering medium with
negative resonance absorption was initially studied in work \cite{Lethohov}
where it was shown that the lasing is possible in such medium. The
scattering by individual active spheres and cylinders was studied
analytically and numerically in a number of works (see, for example \cite
{Kerker}-\cite{Liberal} and references therein). At the same time, the
discussion of the scattering by a cluster of active particles is somewhat
limited to very small clusters made of few particles \cite{Rippol} or to periodic structures \cite{Yannop}.
Scattering by many active particles occurs in so-called random lasers (\cite
{Wiersma}- \cite{Turitsin}), while analytical predictions are difficult to
made for such systems due to extremely large number of particles.

Recently, the scalar wave scattering by dispersive particles was studied by
using the local perturbation method (LPM) in the work \cite{BassVatova}. In
reality, however, the vector wave scattering occurs.

To the author's knowledge, the cluster amplifier made of small particles was
not studied analytically and numerically in the literature before.

In this paper the light amplification and scattering by active particles
with the size smaller than the incident wavelength is studied with the help
of the LPM \cite{Chaumet}-\cite{VP} and the phenomenological laser theory
\cite{Shimoda}-\cite{Haken}. As an example, two important cases are studied
numerically: scattering by ordered cluster and by weakly disordered cluster.
It was shown that the light amplification is possible in both cases, however
it is severely affected by the size of the particles, by the concentration
of the doped active atoms, by interaction between the particles, and by
morphological resonances.

\section{The LPM formalism}

The formalism used in this section is described in a number of papers (\cite
{Chaumet}-\cite{Draine}) and it will be briefly presented here for
convenience and consistency.

Consider the cluster positioned at the origin
of the coordinates and made of $N$ identical active particles which
characteristic size $L$ is small compared to the incident wavelength $%
\lambda $. The frequency-domain fourier transform $\widetilde{\mathbf{E}}(\mathbf{r},\omega)$ of
the electric field $\mathbf{E}(\mathbf{r},t)$ propagating in the host medium
filled with the particles is described by the following equation \cite
{BassVpFr}
\begin{eqnarray}
\left( \bigtriangleup -\mathbf{\nabla }\otimes \mathbf{\nabla }+k^{2}\right)
\widetilde{\mathbf{E}}(\mathbf{r},\omega)+  \notag \\
\frac{k^{2}}{\varepsilon _{h}}\sum_{n=1}^{N}f(\mathbf{%
r}-\mathbf{r}_{n})(\varepsilon _{sc,n}-\varepsilon _{h})\widetilde{\mathbf{E}}(\mathbf{r}%
_{n},\omega)=\mathbf{S}(\mathbf{r}),  \label{plas10}
\end{eqnarray}
where $\mathbf{r}$ and $\mathbf{r}_{n}$ are the radius vectors of the
observer and the $n$-th particle respectively, and
\begin{equation}
k\equiv \frac{2\pi }{\lambda }=\frac{\omega }{c}\sqrt{\varepsilon _{h}},\;f(%
\mathbf{r}-\mathbf{r}_{n})\equiv \left\{
\begin{array}{cc}
1\text{,} & \mathbf{r}\in V_{n} \\
0\text{,} & \mathbf{r}\notin V_{n}
\end{array}
\right. .  \label{plas11}
\end{equation}
Here $\bigtriangleup $ and $\mathbf{\nabla }$\ are the Laplacian and nabla
operators, $\otimes $ defines tensor product, $k$ is a wave number in the
host medium ($\omega $ is the angular frequency and $c$ is the speed of
light in vacuum), $\varepsilon _{sc,n}$ and $\varepsilon _{h}$ are the
relative (in respect to vacuum) permittivities of the $n$-th particle and
the host medium respectively, $f$ is the function describing the shape of
the scatterers, $V_{n}$ is the volume of the $n$-th particle, and $\mathbf{S}
$ is the source of the field. The permittivity $\varepsilon _{sc,n}$\ of the
active particles is, in principle, a complex function depending on the
electric field $\widetilde{\mathbf{E}}(\mathbf{r}_{n},\omega)$ inside the particle, the
frequency $\omega $, and other parameters. We will discuss this topic in
greater detail in the next section.

It should be noted, that the equation (\ref{plas10}) is an approximate one
and it is correct only when the small scatterers ($kL\ll 1$) are considered.

The solution of the equation (\ref{plas10}) can be expressed in the form
\begin{equation}
\widetilde{\mathbf{E}}(\mathbf{r},\omega)\equiv \widetilde{\mathbf{E}}_{in}(\mathbf{r},\omega)+%
\widetilde{\mathbf{E}}_{sc}(\mathbf{r},\omega),  \label{plas12}
\end{equation}
where the scattered field $\widetilde{\mathbf{E}}_{sc}$ is
\begin{equation}
\widetilde{\mathbf{E}}_{sc}(\mathbf{r},\omega)=\frac{k^{2}}{\varepsilon _{h}}\left( \widehat{I}+%
\frac{\mathbf{\nabla }\otimes \mathbf{\nabla }}{k^{2}}\right)
\sum_{n=1}^{N}(\varepsilon _{sc,n}-\varepsilon _{h})\widetilde{\mathbf{E}}(\mathbf{r}%
_{n},\omega)\Phi _{n}(\mathbf{r}),  \label{plas14}
\end{equation}
and
\begin{eqnarray}
\Phi _{n}(\mathbf{r})\equiv \int_{-\infty }^{\infty }\frac{\widetilde{f}(%
\mathbf{q})e^{i\mathbf{q\cdot (r-r}_{n})}}{(q^{2}-k^{2})}d\mathbf{q} \notag \\%
\label{plas18} \\
\widetilde{f}(\mathbf{q})\equiv \frac{1}{8\pi ^{3}}\int_{-\infty }^{\infty
}f(\mathbf{r})e^{-i\mathbf{q\cdot r}}d\mathbf{r.}  \notag
\end{eqnarray}
Here $\widehat{I}$ is the $3\times 3$ unitary tensor in polarization space
and $\mathbf{r}_{n}$ is the radius vector of the n-th particle. The function
$\widetilde{f}$ is the Fourier transform of the function $f$. The incident
field $\widetilde{\mathbf{E}}_{in}$ is created by the source $\mathbf{S}$ in the host
medium (more information can be found in \cite{Jackson}).

The formula (\ref{plas14}) is rather general one and it describes the field
scattered by the cluster made of small particles of arbitrary form. The
resonance properties and interference between the scatterers are taken into
account by the fields $\widetilde{\mathbf{E}}(\mathbf{r}_{n},\omega)$ inside the particles. The
fields $\widetilde{\mathbf{E}}(\mathbf{r}_{n})$ should be found by solving the system of
$3N$ linear equations obtained by substituting $\mathbf{r}=\mathbf{r}_{n}$
into (\ref{plas12}).

The scattered field (\ref{plas14}) can be simplified when the observer is outside of the cluster, such
that $\mathbf{r}\neq \mathbf{r}_{n}$. In this case the integrals (\ref{plas18}) can be
calculated explicitly and the scattered field (\ref{plas14}) can be
presented in the following form
\begin{eqnarray}
\widetilde{\mathbf{E}}_{sc}(\mathbf{r},\omega)=\frac{k^{2}V}{4\pi \varepsilon _{h}}\left(
\widehat{I}+\frac{\mathbf{\nabla }\otimes \mathbf{\nabla }}{k^{2}}\right) \times \notag \\
\sum_{n=1}^{N}(\varepsilon _{sc,n}-\varepsilon _{h})\widetilde{\mathbf{E}}(\mathbf{r}%
_{n},\omega)\frac{e^{ikR_{n}}}{R_{n}},  \label{plas20}
\end{eqnarray}
where
\begin{equation}
R_{n}\equiv \left| \mathbf{r}-\mathbf{r}_{n}\right|, \; \mathbf{r}\neq \mathbf{r}_{n}.  \label{plas21}
\end{equation}
Here $R_{n}$ is the distance between the observation point $\mathbf{r}$ and
the radius vector $\mathbf{r}_{n}$ of the $n$-th scatterer, $V$ is the
scatterer's volume.

In many practical cases the distance between the cluster and the observer is
much larger than the size of the cluster, i.e. $\left| \mathbf{r}\right|
\gg \max (\left| \mathbf{r}_{n}\right| )$, and in addition, the condition $%
k\left| \mathbf{r}\right| \gg 1$ is satisfied. In this case the field (\ref
{plas20}) can be simplified and it can be rewritten in the following form
\begin{eqnarray}
\widetilde{\mathbf{E}}_{sc}(\mathbf{r},\omega)=\frac{k^{2}V}{4\pi \varepsilon _{h}}\frac{e^{ikr}%
}{r}\left( \widehat{I}-\mathbf{l}\otimes \mathbf{l}\right) \times \notag \\
\sum_{n=1}^{N}(\varepsilon _{sc,n}-\varepsilon _{h})\widetilde{\mathbf{E}}(\mathbf{r}%
_{n},\omega)e^{-ik\mathbf{l\cdot r}_{n}},  \label{plas24}
\end{eqnarray}
where
\begin{equation}
\mathbf{l}\equiv \mathbf{r}/r, \; r\equiv \left| \mathbf{r}\right| \gg \max
(\left| \mathbf{r}_{n}\right| ),\text{\ }kr\gg 1.  \label{plas24b}
\end{equation}

We note that the formula (\ref{plas24}) is the final expression for the
field scattered by the cluster of small particles, and it will be used in the following discussion.

\section{The permittivity of the active particles: steady state solution}

In this section we will study the permittivity $\varepsilon _{sc,n}$ of the
active particles which characteristic size is much smaller than the incident
wavelength ($kL\ll 1$). The particles are active due to doped active
atoms with the density $M$. We note that
the permittivity $\varepsilon _{sc,n}$ in formula (\ref{plas24}) can be
complex number with negative or positive
imaginary part, and in this case one can study wave scattering with gain or
loss in active media \cite{Kerker}-\cite{Datsyk}. The problem with such
approach is two-fold. First, the value of the imaginary part is not related
to the properties of the actual medium, and second, the permittivity is the
same for all particles, that is not true for real systems. As it was
suggested in \cite{Karen2}, such approach is valid for quantitative
estimations of lasers before threshold. For more accurate investigation one
should use rigorous methods taking into account atomic transitions and pump
dissipation.

It is important to acknowledge that when the particles are small the number
of active atoms in upper state is constant within the particle, while this
number can be different for other particles. We assume that all the
scatterers have the same density $M$ of the active atoms. It is not limiting
assumption, and it is very convenient one. We also assume that the amplifier
is at steady state, so the densities of the active atoms in the upper
and low states are time independent. We approximate the active atoms as
two-level systems exited by the optical pump with the
frequency $\omega _{p}$ and relaxing with the wide range of
the frequencies $\omega _{j}$ (${j}=1,2,..,p,..,N_{\omega}$). We can present the
permittivity $\varepsilon _{sc,n}$ of the small particles in the following form
\begin{equation}
\varepsilon _{sc,n}\equiv \varepsilon _{sc,n}^{0}+\varepsilon
_{sc,n}^{\prime }(\mathbf{r}_{n},\omega ),  \label{plas26}
\end{equation}
where $\varepsilon _{sc,n}^{0}$ is the permittivity of the $n$-th scatterer
without the active atoms and $\varepsilon _{sc,n}^{\prime }$ is the
permittivity of the $n$-th particle due to the presence of the active atoms.
The latter can be expressed in the following form \cite{Shimoda}
\begin{equation}
\varepsilon _{sc,n}^{\prime }(\mathbf{r}_{n},\omega )=i\frac{c}{\omega }%
\sqrt{\varepsilon _{sc,n}^{0}}\left[ M_{L}(\mathbf{r}_{n})\sigma _{a}(\omega
)-M_{U}(\mathbf{r}_{n})\sigma _{e}(\omega )\right] ,\;  \label{plas26a}
\end{equation}
where the emission and the absorption cross sections of the active medium
respectively are
\begin{eqnarray}
\sigma _{e}(\omega ) &\equiv &\frac{2\pi e^{2}\gamma _{e}\alpha }{mc\sqrt{%
\varepsilon _{sc,n}^{0}}\left( \left( \omega -\Omega _{e}\right) ^{2}+\gamma
_{e}^{2}\right) },  \label{plas26b} \\
\sigma _{a}(\omega ) &\equiv &\frac{2\pi e^{2}\gamma _{a}\alpha }{mc\sqrt{%
\varepsilon _{sc,n}^{0}}\left( \left( \omega -\Omega _{a}\right) ^{2}+\gamma
_{a}^{2}\right) },  \label{plas26c}
\end{eqnarray}
and the density of the active atoms is
\begin{equation}
M=M_{U}(\mathbf{r}_{n})+M_{L}(\mathbf{r}_{n}).  \label{plas26dd}
\end{equation}
Here $M_{U}$ and $M_{L}$ are the densities of the active atoms in the upper
and low states respectively, and $M$ is the total density of the active
atoms. The frequencies $\Omega _{a}$ and $\Omega _{e}$ are the resonance
frequencies for absorption and emission respectively, and $\alpha $ is the
oscillator strength. The frequencies $\gamma _{a}$ and $\gamma _{e}$ are the
dipole relaxation frequencies for the absorption and the emission
respectively (typical values are about $10^{13}$ Hz \cite{Agrawal}), and $e$
and $m$ are the electron's charge and mass respectively.

The importance of the formula (\ref{plas26a}) is that it allows us to use
experimentally measured emission and absorption cross sections ($\sigma _{e}$
and $\sigma _{a}$) when the density $M_{U}$ is known. This approach will be used in the following section where
the results of the numerical calculations will be presented. It should be noted
also that we neglected by the real part of the permittivity $\varepsilon
_{sc,n}^{\prime }$ because it is small compared to the optical contrast $%
\varepsilon _{sc,n}^{0}-\varepsilon _{h}$, especially near the resonance. We
note that the formulae (\ref{plas26b}) were calculated assuming that $\omega
\sim \Omega _{e}$, $\Omega _{a}$.

\subsection{Rate equation approximation}

The expressions (\ref{plas26}) and (\ref{plas26a}) for the permittivity of
the particles suggest that the densities $M_{U}$ and $M_{L}$ should be
known. We can find them by using the rate equation approximation (see for
example, \cite{Shimoda}).

When the pulse duration exceeds the dipole relaxation time (typically $%
10^{-13}$ s), the density $M_{U}$\ of the upper level atoms can be
calculated by using the rate equation approximation in which the dopants
respond so fast that the induced polarization follows the optical field
adiabatically \cite{Agrawal}. We use the following rate equation \cite{Hardy}
\begin{eqnarray}
\frac{dM_{U}(\mathbf{r}_{n})}{dt}=\sum_{j}^{N_{\omega}}I_{j}(\mathbf{r}_{n},\omega ) \times \notag \\
\left[M_{L}(\mathbf{r}_{n})\sigma _{aj}-M_{U}(\mathbf{r}_{n})\sigma _{ej}\right]- %
M_{U}(\mathbf{r}_{n})/\tau ,  \label{plas27}
\end{eqnarray}
where
\begin{eqnarray}
\sigma _{aj} &\equiv &\sigma _{a}(\omega _{j}),\;\sigma _{ej}\equiv \sigma
_{e}(\omega _{j}),  \label{plas28} \\
I_{j}(\mathbf{r}_{n},\omega _{j}) &\equiv &\frac{c\varepsilon _{h}}{4\pi
\hbar \omega _{j}}\left| \widetilde{\mathbf{E}}(\mathbf{r}_{n},\omega _{j})%
\right| ^{2}\Delta \omega ^{2}.  \label{plas28a}
\end{eqnarray}
Here $\sigma _{aj}$ and $\sigma _{ej}$ are the absorption and the emission
cross sections at the frequency $\omega _{j}$ respectively, and $I_{j}$ is
the flux of photons at the frequency $\omega _{j}$, and the $\left| \mathbf{%
...}\right| $ brackets denote the absolute value. The parameter $\tau $ is
the relaxation time of the exited atom (typically $10^{-3}-10^{-6}$ s \cite
{Agrawal}), $\hbar $ is the reduced Planck constant, and $\Delta \omega
=\omega _{j+1}-\omega _{j}$ is the frequency bin.

Since we consider only steady state solutions when
\begin{equation}
\frac{dM_{U}(\mathbf{r}_{n})}{dt}=0,  \label{plas 29}
\end{equation}
the solution of the rate equation (\ref{plas27}) is
\begin{equation}
M_{U}(\mathbf{r}_{n})=\frac{M\sum_{j}^{N_{\omega}}I_{j}(\mathbf{r}_{n},\omega _{j})\sigma
_{aj}}{1/\tau +\sum_{j}^{N_{\omega}}I_{j}(\mathbf{r}_{n},\omega _{j})(\sigma _{aj}+\sigma
_{ej})},  \label{plas31}
\end{equation}
and the formula (\ref{plas26a}) for the permittivity $\varepsilon
_{sc,n}^{\prime }$ has the form
\begin{eqnarray}
\varepsilon _{sc,n}^{\prime }(\mathbf{r}_{n},\omega )\equiv i\frac{cM}{%
\omega }\sqrt{\varepsilon _{sc,n}^{0}} \times \notag \\
\frac{\sigma _{a}(\omega )/\tau
+\sum_{j}^{N_{\omega}}I_{j}(\mathbf{r}_{n},\omega _{j})\left[ \sigma _{a}(\omega )\sigma
_{ej}-\sigma _{e}(\omega )\sigma _{aj}\right] }{1/\tau +\sum_{j}^{N_{\omega}}I_{j}(%
\mathbf{r}_{n},\omega _{j})(\sigma _{aj}+\sigma _{ej})}  \label{plas32}
\end{eqnarray}

The formula (\ref{plas32}) is the main result of this section and it
shows that the permittivity of the small active scatterer is a complex function
of the intensities of the fields $\widetilde{\mathbf{E}}(\mathbf{r}_{n},\omega)
$ inside the particles, frequency $\omega $, position of the particles $%
\mathbf{r}_{n}$, and absorption and emission cross sections $\sigma _{aj}$
and $\sigma _{ej}$. When the light intensity changes (due to increased
reflection from the scatterer's boundaries or due to decreased amount of the
scattered light from all other particles, for example), it will affect the
permittivity $\varepsilon _{sc,n}^{\prime }$ (\ref{plas32}). We note that
the photon fluxes $I_{j}(\mathbf{r}_{n})$ should be found by solving the
system of nonlinear equations with respect to the fields
$\widetilde{\mathbf{E}}(\mathbf{r}_{n},\omega_{j})$ inside the particles.

\subsection{Usage of weak scattering}

As the expression (\ref{plas32}) for the permittivity of the small active particles
suggests, the field scattered by the cluster should be found by solving the
system of nonlinear equations with respect to the fields
$\widetilde{\mathbf{E}}(\mathbf{r}_{n},\omega)$ inside the particles.

However, this tedious task is essentially simplified in our case, because we
consider the scattering by small particles. As the result, the scattered field is
small compared to the incident one, and it is natural to use perturbation
theory where the intensities $I_{j}$ are small compared to the pump intensity $I_{p}$.
We distinct pump and signal (anything but
a pump) frequencies $\omega _{p}$, and $\omega _{j}$ respectively.

When the signal is so small that
\begin{eqnarray}
1/\tau +I_{p}(\mathbf{r}_{n},\omega _{p})(\sigma _{ap}+\sigma _{ep}) &\gg& \notag \\
\sum_{j\neq p}^{N_{\omega}}I_{j}(\mathbf{r}_{n},\omega _{j})(\sigma _{aj}+\sigma _{ej}),
\label{plas35} \\
\sigma _{ap} \equiv \sigma _{a}(\omega _{p}),\;\sigma _{ep}\equiv \sigma
_{e}(\omega _{p}),
\end{eqnarray}
the permittivity (\ref{plas32}) can be expressed in the approximate form
\begin{eqnarray}
\varepsilon _{sc,n}^{\prime }(\mathbf{r}_{n},\omega )=-i\frac{cM}{\omega }%
\sqrt{\varepsilon _{sc,n}^{0}} \times \notag \\
\frac{\sigma _{a}(\omega )/\tau +I_{p}(\mathbf{%
r}_{n},\omega _{p})\left[ \sigma _{a}(\omega )\sigma _{ep}-\sigma
_{e}(\omega )\sigma _{ap}\right] }{1/\tau +I_{p}(\mathbf{r}_{n},\omega
_{p})(\sigma _{ap}+\sigma _{ep})}.  \label{plas37}
\end{eqnarray}
For some estimations it can be sufficient to use the permittivity (\ref
{plas37}), while for rigorous numerical calculations one can apply general
formula (\ref{plas32}) where fluxes $I_{j}$ are found by using the method of
successive approximations.

When emission and absorption spectra are very distinct and separated such
that $\sigma _{ep}=\sigma _{as}=0$, we can simplify the formula (\ref
{plas37}) for the pump and the signal respectively
\begin{eqnarray}
\varepsilon _{sc,n}^{\prime }(\mathbf{r}_{n},\omega _{p}) &=&i\frac{cM}{%
\omega _{p}}\sqrt{\varepsilon _{sc,n}^{0}}\frac{\sigma _{ap}/\tau +I_{s}(%
\mathbf{r}_{n})\sigma _{ap}\sigma _{es}}{1/\tau +I_{p}(\mathbf{r}_{n})\sigma
_{ap}},  \notag \\
&&  \label{plas38} \\
\varepsilon _{sc,n}^{\prime }(\mathbf{r}_{n},\omega _{s}) &=&-i\frac{cM}{%
\omega _{s}}\sqrt{\varepsilon _{sc,n}^{0}}\frac{I_{p}(\mathbf{r}_{n})\sigma
_{es}\sigma _{ap}}{1/\tau +I_{p}(\mathbf{r}_{n})\sigma _{ap}}.  \notag
\end{eqnarray}
The important feature of the formulae (\ref{plas38}) is the sign flip: for
the pump it is positive (the pump is absorbed) and for the signal it is negative
(the signal is amplified).

\section{Intensity of the scattered field and the light amplification}

We define the intensity of the scattered field as $I_{sc}\equiv \left|
E_{sc}\right| ^{2}$, and by using the formula (\ref{plas24}) we can present
the intensity in the following form
\begin{eqnarray}
I_{sc}(\mathbf{r},\omega )=\frac{k^{4}V^{2}}{16\pi ^{2}\varepsilon
_{h}^{2}r^{2}} \times \notag \\
\left| \left( \widehat{I}-\mathbf{l}\otimes \mathbf{l}\right)
\sum_{n=1}^{N}(\varepsilon _{sc,n}-\varepsilon _{h})\widetilde{\mathbf{E}}(\mathbf{r}%
_{n},\omega)e^{-ik\mathbf{l\cdot r}_{n}}\right| ^{2},  \label{plas40}
\end{eqnarray}
where
\begin{equation}
\mathbf{l}\equiv \mathbf{r}/r,\;r\equiv \left| \mathbf{r}\right| \gg \max
(\left| \mathbf{r}_{n}\right| ),\text{\ }kr\gg 1,  \label{plas40aa}
\end{equation}
and the permittivity $\varepsilon _{sc,n}$ is described by the expression (%
\ref{plas32}) or by (\ref{plas37}).

The expression (\ref{plas40}) suggests that the light amplification (related
to the imaginary part of the permittivity $\varepsilon _{sc,n}$) is due to
step wise amplification inside each active particle, and it is coded in the fields
$\widetilde{\mathbf{E}}(\mathbf{r}_{n},\omega)$. Below we consider the
fields $\widetilde{\mathbf{E}}(\mathbf{r}_{n},\omega)$ in grater
detail for one active particle.

\begin{figure}[t]
\begin{center}
\includegraphics [width=9.1cm]
{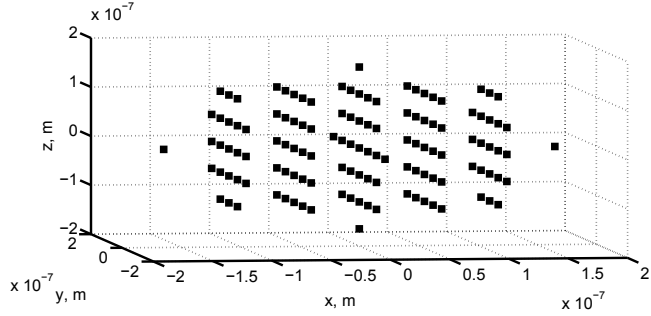}
\end{center}
\caption{The schematic representation of the spherical cluster made of 123 small cubes. The period of the cluster is d=2.2L,
and the characteristic size of the cubes is $L=25$ nm.}
\label{fig1}
\end{figure}

\subsection{The light amplification by small active sphere}

Consider the light amplification by small active sphere. In this case the
intensity of the scattered field is described by the expression (\ref{plas40}%
) where the field $\widetilde{\mathbf{E}}(\mathbf{r}_{1},\omega)$ has the following form \cite
{BassFRVP}
\begin{equation}
\widetilde{\mathbf{E}}(\mathbf{r}_{1},\omega )=\frac{\widetilde{\mathbf{E}}_{in}(\mathbf{r}%
_{1},\omega )}{D(\omega )},  \label{plas401}
\end{equation}
and the denominator $D$ is
\begin{equation}
D(\omega )=1+\frac{(\varepsilon _{sc,1}-\varepsilon _{h})}{3\varepsilon _{h}}%
\left( 1-L^{2}k^{2}-i\frac{2}{3}L^{3}k^{3}\right) .  \label{plas402}
\end{equation}
The resonance frequency $\omega _{r}$ is found from the following equation
\begin{equation}
\operatorname{Re}D(\omega _{r})=0,  \label{plas403}
\end{equation}
and the resonance width $\xi $ is defined as
\begin{equation}
\xi \equiv\left| \frac{\operatorname{Im}D(\omega)}{\frac{\partial \operatorname{Re}%
D(\omega )}{\partial \omega }}\right|_{\omega=\omega_{r}} .
\label{plas404}
\end{equation}
In accordance with formulae (\ref{plas403}) and (\ref{plas404}) the
resonance width $\xi$ and the resonance frequency $\omega_{r}$ respectively are
\begin{equation}
\xi =\frac{\sqrt{\varepsilon _{h}}L\omega _{r}^{2}}{3c}+\frac{3c^{2}\operatorname{Im}%
(\varepsilon _{sc,1}-\varepsilon_{h})}{2\omega _{r}L^{2}\operatorname{Re}^{2}(\varepsilon
_{sc,1}-\varepsilon _{h})},  \label{plas405}
\end{equation}
and
\begin{equation}
\omega _{r}=\frac{c\sqrt{3}}{L\sqrt{\operatorname{Re}(\varepsilon
_{sc,1}-\varepsilon _{h})}}\left( 1+\frac{\operatorname{Re}(\varepsilon
_{sc,1}-\varepsilon _{h})}{3\varepsilon _{h}}\right) ^{1/2}.  \label{plas408}
\end{equation}
When the permittivity $\varepsilon _{sc,1}$ of the particle is real, the
resonance width is defined by the first term in Eq. (\ref{plas405}). When
the imaginary part of the permittivity $\varepsilon _{sc,1}$ is taken into
account and it is negative or positive, the resonance width can be slightly
decreased or increased respectively. The formulae (\ref{plas405}) and (\ref{plas26a})
suggest that when $M_{U}\sigma _{e}<M_{L}\sigma _{a}$, the resonance
width increases (with respect to the one in the passive medium) and it
decreases when $M_{U}\sigma_{e}>M_{L}\sigma _{a}$.

This decrease (or increase) corresponds to effective gain (or loss) of the field scattered by the
particle at the resonance frequency.

 We note that similar conclusions can be drawn for the clusters consisting of two and more particles,
 while the analytical investigation of such systems is much more complicated.

\begin{figure}[t]
\begin{center}
\includegraphics [width=9.3cm]
{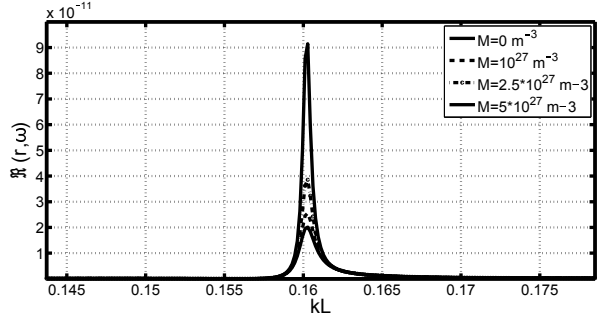}
\end{center}
\caption{The normalized intensity of the scattered field $\Re $ versus
normalized frequency $kL$ for the spherical cluster made of small cubes with
different density of the active atoms $M$. The period of the cluster is d=2.2L, and
the permittivity of the particles and the host medium is $\protect\varepsilon _{sc,n}=4.2466$ and $\protect%
\varepsilon _{h}=1$ respectively, the characteristic size of the cubes is $%
L=25$ nm, and the total number of the particles in the cluster is $N=515$.}
\label{fig2}
\end{figure}

\section{Two numerical examples: light amplification by ordered and by weakly
disordered spherical cluster}

In this section the light amplification and scattering by active clusters (clusters made of an active material)
is studied numerically. The normalized intensity $\Re $
of the scattered field is calculated. The normalized intensity is defined as
\begin{equation}
\Re (\mathbf{r},\omega )\equiv I_{sc}(\mathbf{r},\omega )/I_{inc}(0,\omega ).
\end{equation}

All the used clusters are 3D structures (as shown, for example, on the figure \ref{fig1}) made
of cubes doped with $Yb^{3+}$ (active material). The absorption and emission cross
sections are taken from \cite{Pask} and the other parameters are from \cite{Hardy}.
The incident field is generated by the point source described by the following formula
\begin{equation}
\widetilde{\mathbf{E}}_{in}(\mathbf{r},\omega )=\mathbf{E}_{0}\frac{e^{ik(\mathbf{r-r}%
_{s})}}{4\pi \left| \mathbf{r-r}_{s}\right| },\;k\left| \mathbf{r-r}%
_{s}\right| \gg 1,
\end{equation}
where the field $\mathbf{E}_{0}$ is polarized along $z$ direction. The
source is positioned at $\mathbf{r}_{s}=\{1,0,0\}$ and the center of the
cluster is positioned at the origin of coordinates $\mathbf{r}=\{0,0,0\}$.
The pump wavelength is selected to be $\lambda_{p}=911$ nm, and at this specific wavelength the field $%
\mathbf{E}_{0}$ is artificially increased by several orders of magnitude to simulate the pump.
Finally, the perturbation theory is used to calculate the density of the
upper level atoms.

\begin{figure}[t]
\begin{center}
\includegraphics [width=9.3cm]
{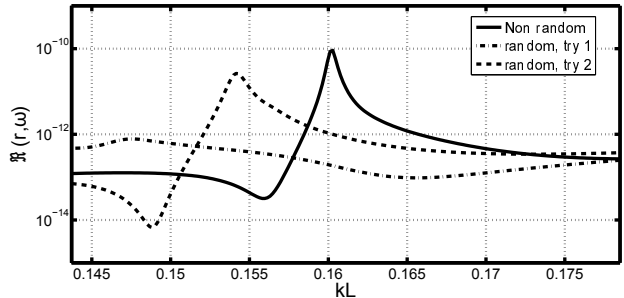}
\end{center}
\caption{The normalized intensity of the scattered field $\Re $ versus
normalized frequency $kL$ for slightly disordered spherical cluster made of
small cubes. The results of two runs are presented, and the scattering by
the ordered cluster is also shown for comparison. The period of the cluster is d=2.2L, and
the density of the active atoms is $M=5\ast 10^{27}$ m$^{-3}$.
The permittivity of the particles and the
host medium is $\protect\varepsilon _{sc,n}=4.2466$ and $\protect\varepsilon %
_{h}=1$ respectively, the characteristic size of the cubes is $L=25$ nm, and
the total number of the particles in the cluster is $N=515$.}
\label{fig3}
\end{figure}

\subsection{The amplification and scattering by ordered spherical cluster}

Consider the scattering by the spherical cluster made of the small active
cubes organized into simple cubic lattice. The size of the cubes in the
cluster is $L=25$ nm, the period is $d=2.2L$, the radius
of the cluster is $5d$. The permittivity of the scatterers is $\varepsilon
_{sc,n}=4.2466$ and the permittivity of the host medium is $\varepsilon _{h}=1$.
Such combination of the cluster's permittivity and dimensions creates optical
resonance of the passive cluster (cluster without any active material)
near $kL=0.16$ ($\lambda =982$ nm).

The figure \ref{fig2} shows the normalized intensity $\Re $ of the scattered
field for the active clusters with $Yb^{3+}$ density $M=5\ast 10^{27}$ m$^{-3}$
(upper solid line), $M=2.5\ast 10^{27}$ m$^{-3}$ (dash-dotted line), and $M=10^{27}$
m$^{-3}$ (dashed line). For comparison, the intensity of the scattered
field from the passive cluster with $Yb^{3+}$ density $M=0$ m$^{-3}$ (lower solid line)
is also presented.

The figure \ref{fig2} shows that light is significantly amplified at the selected frequency $%
kL=0.16$ for the doping exceeding $M=10^{27}$ m$^{-3}$.
For relatively low $M$, when the doping increases 2.5 times
(from $M=10^{27}$ m$^{-3}$ to $M=2.5\ast10^{27}$ m$^{-3}$), the
intensity grows only 1.6 times. However, for relatively higher $M$, when the
doping increases only 2 times (from $M=2.5\ast10^{27}$ m$^{-3}$ to $M=5\ast10^{27}$ m$^{-3}$),
the intensity grows 2.25 times. Additional simulations (not presented here) suggest that at even high doping,
the light amplification increases several orders of magnitude while
the doping increases only few times.

We realize that the size of the cluster used in our calculation is too small to made a lasing with the
conventional values of the doping $M$ ($M\sim 10^{25}$ m$^{-3}$) and
that is why we have presented only results with $M$ smaller than physically realistic limit ($\sim
10^{28}$ m$^{-3}$).

\subsection{The amplification and scattering by spherical cluster with weak
positional disorder}

In this subsection we consider the light amplification and scattering by the
active cluster which particles are randomly positioned near predefined positions.
The predefined positions are the nodes of the cubic lattice with the period
of $d=2.2L$, and the particles are positioned not further than $0.1L$ from
the nodes to avoid a collision. We note that the distance $0.1L$ is actually
$2.5$ nm, that is much less than the size of the particle $L$, and that is
why we call this cluster weakly disordered one.

We note that the density of the active atoms in the cluster is $M=5\ast 10^{27}$ m$^{-3}$.
The permittivity of the particles and the host medium is
$\protect\varepsilon _{sc,n}=4.2466$ and $\protect\varepsilon %
_{h}=1$ respectively, the characteristic size of the cubes is $L=25$ nm, and
the total number of the particles in the cluster is $N=515$.

The results of the calculations are presented on the Figure \ref{fig3}. The
figure shows the results of two simulation runs for the disordered cluster (dashed and dash
dotted lines) and one result for ordered cluster (solid line) for comparison.

The figure suggests that despite weak positional disorder the random
positioning significantly influences the scattering and amplification by the
cluster. In our particular case, two scattering peaks compete with each
other: one near $kL=0.154$ ($\lambda =1020$ nm) and another is near $kL=0.147$
($\lambda =1069$ nm). This feature is probably related to the emission spectrum
of active material ($Yb^{3+}$) which has two crests: very narrow one near $%
980$ nm and broad one near $1030$ nm.

The random amplifier differs from the nonrandom one in a number of ways. The
first difference is the absence of well defined boundaries, which in turn,
govern the morphological resonances. Thus, the amplification (or lasing) can
be at several frequencies simultaneously. The second difference is the
random structure inside the cluster, affecting the interaction between the
particles and the total gain as the result.

\section{Conclusions}

The light amplification and scattering by the cluster made of the small active
particles have been studied analytically and numerically.

The permittivity of the small active particles has been calculated in steady state
by using the rate equation approximation.

The light amplification has been discussed for small active particle. It has been suggested
that the amplification (or loss) effectively occurs in the active particle due to
narrowing (or broadening) of the resonance.

The light scattering by the ordered and slightly disordered clusters of small active particles has been
calculated numerically. The numerical simulations have been shown that the light amplification occurs near
the morphological resonances which are governed by the shape of the cluster and its optical contrast.

\begin{equation*}
\end{equation*}

\textbf{Acknowledgments}

Many thanks to my wife Lucy for encouragement, understanding,\ and support.

\begin{equation*}
\end{equation*}

\end{document}